\documentclass[aps,amsmath,showpacs,showkeys,superscriptaddress,twocolumn,10pt]{revtex4-2}
\usepackage{color}
\usepackage{amssymb}
\usepackage{titlesec}
\usepackage{amssymb,amsfonts,amsthm}
\begin{document}
\title{Schwarzian derivative in higher-order Riccati equations}
\author{Benoy Talukdar}
\affiliation{Department of Physics, Visva-Bharati University, Santiniketan 731235, India}
\author{Supriya Chatterjee}
\affiliation{Department of Physics, Bidhannagar College, EB-2, Sector-1, Salt Lake, Kolkata 700064, India}
\author{Golam Ali Sekh} 
\email{golamali.sekh@knu.ac.in}
\affiliation{Department of Physics, Kazi Nazrul University, Asansol 713340, India}

\begin{abstract}
The Sturm-Liouville equation represents the linearized form of the first-order Riccati equation. This provides an evidence for the connection between Schwarzian derivative and this first-order nonlinear differential equation. Similar connection is not obvious for higher-order equations in the Riccati chain because the corresponding linear equations are of order greater than two. With special attention to the second- and third-order Riccati equations we demonstrate that Schwarzian derivative
has a natural space in higher Riccati equations. There exist higher-order analogues of the Schwarztan derivative. We demonstrate that equations in the Riccati hierarchy are embedded in these higher-order derivatives.
\end{abstract}
\maketitle
 %%%%%%%%%%%%%%%%%%%%%%%%%%%%%%%%%%%%%%%%%%%%%%%%%%%%%%%%%%%%%%%%%%%%%%%
\section*{1. Introduction}
Let $f(x)$ represent a well behaved function of a single variable $x$ (real or complex) and primes over it denote appropriate derivatives,  then $S(f(x))$ written as
\begin{equation}
 S(f(x))=\frac{f'''(x)}{f'(x)}-\frac{3}{2}(\frac{f''(x)}{f'(x)})^2
\end{equation}
is called the Schwarzian derivative of $f(x)$ or Schwarzian  in short \cite{1,2}. The expression in (1) appears in a wide variety of mathematical and physical context ranging from classical complex analysis to conformal field theory \cite{3}.
\par In the theory of complex variables the linear fractional or M\"{o}bius transformation given by
\begin{equation}
 f(x)=\frac{ax+b}{cx+d}
\end{equation}
posses a number of remarkable properties. In particular, the transformation (2) with real coefficients forms  a group of  symmetries in real projective space \cite{4,5}. The Schwarzian derivative of $f(x)$ in (2) is zero. Thus for any arbitrary function, $S(f(x))$ measures how much $f$ differs from being a M\"{o}bius transformation. It is important to note that the Schwarzian is not a function but represents the difference between a function and a more complicated tensor field \cite{6}.

 An important property of the Swarzian is that $S(f)=S(g)$ if and only if $g(x)=(af(x)+b)/(cf(x)+d)$ with $ad-bc\neq0$. It is interesting to note that if $\psi_1(x)$ and $\psi_2(x)$ represent two linearly independent solutions of the general Sturm-Liouville equation 
\begin{subequations}
 \begin{equation}
 \psi''(x)+p(x)\psi'(x)+q(x)\psi(x)=0
\end{equation}
\mbox{and}
\begin{equation}
 f(x)=\frac{\psi_1(x)}{\psi_2(x)},
\end{equation}
\end{subequations}
then
\begin{equation}
 S(f(x))=-\frac{1}{2}p(x)^2+2q(x)-p'(x).
\end{equation}
For $p(x)=0$, (3a) corresponds to the time-independent Schr\"{o}dinger equation with $q(x)$ as the potential function.  Using $p(x)=p'(x)=0$ in (4) we see that the Schwarzian reconstructs the potential. Assuming 
\begin{equation}
 \psi(x)=u(x)v(x)
\end{equation}
as a complete solution of (3) it is straightforward to write the equation in the normal form
\begin{equation}
 v''(x)+\lambda(x)v(x)=0
\end{equation}
with $\lambda(x)=S(f(x))/2$. This shows that when the first derivative is removed from the Sturm-Liouville equation the coefficient of the dependent variable is equal to half the Schwarzian derivative. Apparently, this simple fact has remained un-noticed in the literature.
%Apparently, this simple fact has remained un-noticed in the literature. The object of the present  paper is to exploit this fact to demonstrate the appearance of Schwarzian derivative in higher-order  equations in the Riccati chain \cite{7} rather than  following the method applicable for the first-order equation. In this context we shall see that linearization of the higher-order Riccati equation leads to differential equations of order greater than 2. This poses an awkward analytical constraint to look for the appearance of $S(f(x))$ in these equations.
One of our objectives in this work is to exploit this fact to demonstrate the appearance of Schwarzian derivative in higher-order equations in the Riccati chain\cite{7}. As regards our other point of  interest, we show that higher-order Schwarzian derivatives\cite{8a} lead, in a rather natural way, to a hierarchy of nonlinear differential equations and examine if these equations could be converted to the Riccati form.

In section 2 we introduce the equations in the Ricatti chain and point out that each equation in the chain can be reduced to the linear form. We then show that the set of linear equations so obtained provides a basis to visualize how the Schwarzian derivative enters into the Ricatti equations. In section 3 we introduce the higher-order Schwarzian derivatives for normalized univalent functions as considered in ref. \cite{8a}. In a relatively recent publication \cite{9a}, the first-order Ricatti equation was realized in the context of a Schwarzian equation provided by the derivative as introduced through Eq.(1). In the same way, we  obtain nonlinear equations associated with higher-order Schwarzian derivatives and suggest a method to reduce theme to the Riccati form. Finally, in section $4$ we summarize our outlook on this work and try to make some concluding remarks.

\section*{2. Riccati equations and Schwarzian}
The Riccati equation  represented  by \cite{8}
\begin{equation}
 \frac{d\omega(x)}{dx}+\alpha(x)+\beta(x)\omega(x)+\omega(x)^2=0
\end{equation}
is the only nonlinear differential equation that is of Painleve type \cite{9}. It appears in many different fields of  physics and mathematics \cite{10,11,12} and can be linearized using  the Cole-Hopf transformation \cite{13,14}
\begin{equation}
\omega(x)=\frac{\phi'}{\phi}
\end{equation}
to write
\begin{equation}
 \phi''(x)+\beta(x)\phi'(x)+\alpha(x)\phi(x)=0.
\end{equation}
Thus the Schwarzian derivative of (9) will be of the same form as that in (4) with $p(x)=\beta(x)$ and $q(x)=\alpha(x)$. As expected, if (9) is expressed in the normal form removing the first derivative by the use of $\phi(x)=u(x)v(x)$ then the                                                                                            coefficient of $v(x)$ will be half of the Schwarzian  derivative of (9).

Following Grundland and Levi \cite{7} we write the $n$-th order equation  in the  Riccati chain as
\begin{equation}
 L^n\omega(x)+\sum_{i=1}^n\alpha_i(x)(L^{i-1}\omega(x))+\alpha_0(x)=0.
\end{equation}
Here $n$ denotes the order of the Riccati equation and $L$ stands for a differential operator represented by
\begin{equation}
 L=\frac{d}{dx}+c\omega(x).
\end{equation}
The transformation in (8) re-written as $\omega(x)=\phi'(x)/c\phi(x)$ converts the equation  in (10) into a linear ordinary differential given by
\begin{equation}
 \sum_{i=0}^n\alpha_i(x)\frac{d^i\phi(x)}{dx^i}+\frac{d^{n+1}\phi}{dx^{n+1}}=0.
\end{equation}
The second-order Riccati equation and corresponding linear equation obtained by using $n=2$ in (10) and (12) are
\begin{equation}
\begin{split}
 \omega''(x)+(3cf(x)+\alpha_2(x))\omega'(x)+c\alpha_2(x)\omega(x)^2\\+c^2\omega(x)^3+\alpha_1(x)\omega(x)+\alpha_0(x)=0
\end{split}
\end{equation}
and
\begin{equation}
 \phi^{(3)}(x)+\alpha_2(x)\phi''(x)+\alpha_1(x)\phi'(x)+c\alpha_0(x)\phi(x)=0.
\end{equation}
The space of solutions of (14) is three dimensional spanned by three linearly independent eigen functions, say, $\phi_1(x)$, $\phi_2(x)$ and $\phi_3(x)$. This tends to pose some problem to construct the Schwarzian corresponding to (14). Clearly, the same is true for other equations in the hierarchy. However, the appearance of the Schwarzian derivative in the second-order Riccati equation can be seen as follows. 

Making use of $\phi(x)=u(x)v(x)$ in (14) we get
\begin{equation}
 v^{(3)}(x)+a_2^2(x)v''(x)+a_3^2(x)v'(x)+a_4^2(x)v(x)=0,
\end{equation}
where
\begin{subequations}
 \begin{equation}
 a_2^2(x)=\alpha_2(x)+\frac{3u'(x)}{u(x)},
 \end{equation} 
\begin{equation}
 a_3^2(x)=\alpha_1(x)+\frac{2\alpha_2(x)u'(x)}{u(x)}+\frac{3u''(x)}{u(x)}
\end{equation}
\mbox{and}
\begin{equation}
a_4^2(x)=c\alpha_0(x)+\frac{\alpha_1(x)u'(x)}{u(x)}+\frac{\alpha_2(x)u''(x)}{u(x)}+\frac{u^{(3)}(x)}{u(x)} 
\end{equation}
\end{subequations}
We now choose $u(x)$ such that $a_2^2(x)=0$. This gives 
\begin{equation}
 u(x)=e^{-\frac{1}{3}\int\alpha_2(x)dx}.
\end{equation}
From (15) and (17) we find
\begin{equation}
 v^{(3)}(x)+b_3^2(x)v'(x)+b_4^2(x)v(x)=0
\end{equation}
with
\begin{subequations}
 \begin{equation}
 b_3^2=\alpha_1(x)-\frac{\alpha_2(x)^2}{3}-\alpha_2'(x)
\end{equation}
\mbox{and}
\begin{equation}
 b_4^2=c\alpha_0(x)-\frac{\alpha_1(x)\alpha_2(x)^2}{3}-\frac{2\alpha_2(x)^2}{27}-\frac{\alpha_2''}{3}.
\end{equation}
\end{subequations}
Equation (18) represents a third-order linear differential equation with the second derivative removed and this result has been found from the second-order Riccati equation. It is easy to see that the coefficient $b_3^2$ of $v'(x)$ in (18) stands  for the Schwarzian derivative of the Sturm-Liouvlle equation (3) with $p(x)=2\alpha_2(x)/3$ and $q(x)=\alpha_1(x)/3$. This is how the Schwarzian  derivative appears in the second-order Riccati equation.
\par The expression for $b_4^2$, the coefficient of $v(x)$ in (18) involves the second-order derivative of $\alpha_2(x)$ and thus it appears that $b_4^2$ is unlikely to have any connection with Schwarzian. We, however found that the combination $(b_4^2(x)-b_3'^2(x)/3)/\alpha_2(x)$ under the constraint $\alpha_1'(x)=3c\alpha_0(x)$ represents a Schwarzian of (3) with $p(x)=2\alpha_2(x)/3$ and $q(x)=\alpha_2(x)/2$.
\par The third-order Riccati equation is given by
\begin{equation}
\begin{split}
\omega^{(3)}(x)+(4c\omega(x)+\alpha_3(x))\omega''(x)+(6c^2\omega(x)^2\\+3c\alpha_3(x)\omega(x)+\alpha_2(x))\omega'(x)+3c\omega'(x)^2\\+c^3\omega(x)^4+c^2\alpha_3(x)\omega(x)^3\\+c\alpha_2(x)\omega(x)^2+\alpha_1(x)\omega(x)+\alpha_0(x)=0.
\end{split}
\end{equation}
The corresponding linear equation reads
\begin{equation}
\begin{split}
 \phi^{(4)}(x)+\alpha_3(x)\phi^{(3)}(x)+\alpha_2(x)\phi''(x)\\+\alpha_1(x)\phi'(x)+c\alpha_0(x)=0.
\end{split}
\end{equation}
When the third derivative term in (21) is removed we get
\begin{equation}
 v^{(4)}(x)+b_3^3(x)v''(x)+b_4^3(x)v'(x)+b_5^3(x)v(x)=0
\end{equation}
with
\begin{subequations}
 \begin{equation}
 b_3^3(x)=\alpha_2(x)-\frac{3\alpha_3(x)^2}{8}-\frac{3\alpha_3'(x)}{2}, 
 \end{equation}
\begin{equation}
 b_4^3(x)=\alpha_1(x)-\frac{\alpha_2(x)\alpha_3(x)}{2}+\frac{\alpha_3(x)^3}{8}-\alpha_3''(x) 
\end{equation}
\mbox{and}
\begin{equation}
 \begin{split}
 b_5^3(x)=c\alpha_0(x)-\frac{\alpha_1(x)\alpha_3(x)}{4}+\frac{\alpha_2(x)\alpha_3(x)^2}{16}-\frac{3\alpha_3(x)^4}{256}\\-\frac{\alpha_2(x)\alpha_3'(x)}{4}+\frac{3\alpha_3(x)^2\alpha_3'(x)}{32}+\frac{3\alpha_3'(x)^2}{16}-\frac{\alpha_3(x)^3}{4}.
 \end{split}
\end{equation}
\end{subequations}
The expression for the coefficient $b_3^3(x)$ of $v''(x)$ in (22) involves only the first derivative of $\alpha_3(x)$. Thus , as in the case of second Riccati equation, the result for $b_3^3(x)$ represents a Schwarzian derivative corresponding to (3) with $p(x)=\alpha_3(x)/2$ and $q(x)=\alpha_2(x)/6$.
From $b_3^3(x)$ and $b_4^3(x)$ we have verified that under the constraint $\alpha_2'(x)=3\alpha_1(x)/2$ the quantity $(b_4^3(x)-2b_3'^2(x))/\alpha_3(x)$ is a Schwarzian $-\alpha_2(x)+\alpha_3(x)^2/4+\alpha_3'(x)$. The method applied for the second- and third-order equations can easily be adapted to look for the presence of Schwarzian derivative in still higher-order Riccati equations.

\section*{3.Ricatti equations from higher-order Schwarzians}
 The higher-order-Schwarzian derivatives are defined inductively \cite{8a} by
\begin{eqnarray}
\sigma_{n+1}(f)=\sigma(f)'-(n-1)\frac{f''}{f}\sigma_n(f),\,\,n\geq 3.
\label{eq24}
\end{eqnarray}
with $\sigma_3(f)=S(f(x))$ as given in (1). From (\ref{eq24}) we can easily deduce
\begin{subequations}
\begin{eqnarray}
\sigma_4(f)=\frac{f^{(4)}(x)}{f'(x)}-6\frac{f''' f''}{{f'}^2}+6\left(\frac{f''}{f'}\right)^3,
\end{eqnarray}
\begin{eqnarray}
\sigma_5(f)&=&\frac{f^{(5)}}{f'}-10\frac{f^{(4)f''}}{f'^2}-6\left(\frac{f'''}{f'}\right)^2\nonumber\\&+&48\frac{f'''f''^2}{f'^3}-36\left(\frac{f''}{f'}\right)^4,
\end{eqnarray}
\begin{eqnarray}
\sigma_6(f)&=&\frac{f^{(6)}}{f'}-15\frac{f''f^{(5)}}{f'^2}-22\frac{f'''f^{(4)}}{f'^2}\nonumber\\&+&108\frac{f''^2f^{(4)}}{f'^3}+132\frac{f''(f^{(3)})^2}{f'^4}\nonumber\\&-&480\frac{f''^3f^{(3)}}{f'^ 4}+288\frac{f''^5}{f'^5},
\end{eqnarray}
\end{subequations}
and similar results for still higher-order Schwarzian derivatives.

If we denote the  pre-Schwarzian derivative $f''(x)/f'(x)$ by $y(x)$ the Schwarzian equation \cite{r17}
\begin{eqnarray}
S(f(x))=-g(x)
\label{eq26}
\end{eqnarray}
with $g(x)$  as an external source function leads to the nonlinear equation
\begin{eqnarray}
y'(x)+\frac{1}{2} y(x)^2-g(x)=0.
\label{eq27}
\end{eqnarray}
Equation (\ref{eq27}) was identified as Ricatti equation  presumably because it could be linearized by Cole-Hopf transformation \cite{13,14}.

We feel that it will be  interesting to examine if  equations similar to that in (\ref{eq26}) written using higher-order Schwarzians  lead to higher-order Ricatti equations . To that end we first begin with (25a) and write the nonlinear equation for the pre-Schwarzian derivative $y(x)$ as
\begin{eqnarray}
y''(x)-3y'(x)y(x)+3 y(x)^3+g_1(x)=0.
\label{eq28}
\end{eqnarray}
It is straightforward to see that the Cole-Hopf transformation  $y(x)=-\psi'(x)/\psi(x)$ reduces  (\ref{eq28}) in the linear form such that this nonlinear equation is, in fact, the second-order Ricatti equation. In respect of this the nonlinear equations constructed by using $\sigma_i(f)$ for $i\geq 4$ tend to pose problem. In the following we demonstrate this by dealing with $\sigma_5(f)$.

The relation in (25b) leads to the nonlinear equation
\begin{eqnarray}
y^{(3)}(x)&-&6 y''(x) y(x)-3y'(x)^2+12 y'(x)y(x)^2\nonumber\\&-&3y(x)^4+g_2(x)=0.
\label{eq29}
\end{eqnarray}
The transformation used for linearizing (\ref{eq28}) reduces (\ref{eq29}) to the form
\begin{eqnarray}
\frac{\psi^{(4)}(x)}{\psi(x)}+\frac{2\psi'(x)\psi^{(3)}(x)}{\psi(x)^2}-g_2(x)=0.
\end{eqnarray}
Thus we infer that the nonlinear equation (29)  when appended by $-2\psi'(x)\psi^{(3)}(x)/\psi^2(x)$  (expressed in terms of $y(x)$ and derivatives of $y(x)$) will give the third equation in the Riccati  chain. Similarly, the fourth-order Riccati  equation can be obtained from the nonlinear equation (following from $\sigma_6(f)$ )
\begin{eqnarray}
y^{(4)}(x)&-&10 y(x)y^{(3)}(x)-12 y'(x) y''(x)\nonumber\\ &+&36 y(x)^2y''(x)+36 y(x)y'(x)^2\nonumber\\&-&60 y'(x)y(x)^3+12y(x)^5+g_3(x)=0
\label{eq31}
\end{eqnarray} 
 by adding $-4\psi'(x)^2\psi^{(3)}(x)/\psi(x)^3-2\psi'(x)(\psi^{(3)}(x))^2-5\psi'(x)\psi^{(4)}(x)/\psi(x)^2$ to it.
 
 The higher-order Schwarzian derivatives as used in this work were proposed by Schippers \cite{8a} to  fit   the flow that satisfies the 
 Loewner differential equation \cite{r18}. In addition, Aharanov \cite{r19} and Tamanoi \cite{r20} gave two different definitions to introduce the higher-order analogues of the Schwarzian derivative. In particular. Aharanov  gave the necessary and sufficient condition for a non-constant meromorphic function on the unit disc to be univalent in terms of his Schwarzian, on the other hand,  Tamanoi studied combinatorial structures of his Schwarzians. We have checked and verified that derivatives provided by the Scippers’ scheme are more suitable for the present study. 
 
 In the above we have seen that he Schwarzian equations formed by using $\sigma_n$  for $n\geq 5$  tend to exhibit some deviation from the corresponding Riccati equations. However, one can choose appropriate weight factors for the numerical coefficients in  higher-order Schwarzian equations to reduce them in the Riccati form. For example, such reduced equations corresponding to (\ref{eq29}) and (\ref{eq31}) are given by
\begin{eqnarray}
 y^{(3)}(x)&-&4 y(x) y''(x)-3 y'^2(x)+6y(x)^2y'(x)\nonumber\\&-&y(x)^4+g_2(x)=0
 \label{eq32}
\end{eqnarray}
\begin{eqnarray}
y^{(4)}(x)&-&5 y(x)y^{(3)}(x)-10y'(x)y''(x)\nonumber\\&+&10y(x)^2y''(x)+15y(x)y'(x)^2\nonumber\\&-&10 y(x)^3 y'(x)+y(x)^5+g_3(x)=0.
\label{eq33}
\end{eqnarray}
The equations (\ref{eq32}) and (\ref{eq33}) can be linearized by using $y(x)=-\psi'(x)/\psi(x)$ to read
\begin{eqnarray}
&&\psi^{(4)}(x)-g_2(x)\psi(x)=0\nonumber\\{\rm and} &&\,\,\psi^{(5)}(x)-g_3(x)\psi(x)=0
\label{eq34}
\end{eqnarray}
respectively such that (\ref{eq32}) and (\ref{eq33}) indeed represent the third and fourth equations in the Riccati chain.

\section*{4. Conclusion}
%In the recent past Carillo et \cite{15} used the properties of Schwarzian derivative to obtain solutions of the second-order  Riccati equation. 
In the present work we examined the connection between the Schwarzian derivative and higher-order equations in the Riccati chain. We achieved this by removing the $n$-th order derivative term from the $(n+1)$-th order linear  differential equation associated with $n$-th order Riccati equation. About fifty years ago such a reduced equation was considered  by Kim \cite{16} in order to look for some  generalization of the Schwarzian derivative. We note that the generalized derivatives so obtained lead to highly complicated expressions for the coefficients of first- and zeroth-order derivative terms of a third-order equation. As opposed to this we derived a transparent method  to demonstrate how the so-called traditional Schwarzian derivative enters into the higher-order equations of the Riccati Chain.

In the above context we also noted that there are three well known methods \cite{8a,r18,r19} to introduce higher-order Schwarzian derivatives. But we found it convenient to work with the definition of Schippers [8] and thus provided an approach to realize the equations in the Riccati chain using inductively generated hierarchy of  higher-order Schwarzian derivatives. 
%One may, however, improve on the method followed by us.

\noindent{\bf Acknowledgement}\\
One of the authors (GAS) would like to acknowledge funding from the `Science and Engineering Research Board, Govt. of India' through Grant No. CRG/2019/000737.

\end{document}